\documentclass[a4paper,11pt]{article}
\usepackage{epsfig}
\usepackage{url}
\usepackage{graphicx,color}
\usepackage[psamsfonts]{amssymb}
\usepackage{amsmath}
\usepackage{indentfirst}
\usepackage{mathrsfs}
\usepackage{booktabs}
\usepackage{tabularx}
\usepackage{authblk}
\usepackage{setspace}

\usepackage{color}

\title{The Extreme Risk of Personal Data Breaches\\ \& \\ The Erosion of Privacy}
\author{Spencer Wheatley$^1$, Thomas Maillart$^2$ and Didier Sornette$^1$\\
$^1$ {\small ETH Zurich, Department of Management, Technology and Economics, Switzerland}\\
$^2$ {\small School of Information, University of California, Berkeley, 102 South Hall, Berkeley, CA 94720, USA}\\
{\small e-mails: swheatley@ethz.ch, maillart@ischool.berkeley.edu and dsornette@ethz.ch}
}

\begin{document}

\maketitle

\begin{abstract}

Personal data breaches from organisations, enabling mass identity fraud, constitute an \emph{extreme risk}. This risk worsens daily as an ever-growing amount of personal data are stored by organisations and on-line, and the \emph{attack surface} surrounding this data becomes larger and harder to secure. Further, breached information is distributed and accumulates in the hands of cyber criminals, thus driving a cumulative erosion of privacy. Statistical modeling of breach data from 2000 through 2015 provides insights into this risk: A current maximum breach size of about 200 million is detected, and is expected to grow by fifty percent over the next five years. The breach sizes are found to be well modeled by an \emph{extremely heavy tailed} truncated Pareto distribution, with tail exponent parameter decreasing linearly from 0.57 in 2007 to 0.37 in 2015. With this current model, given a breach contains above fifty thousand items, there is a ten percent probability of exceeding ten million. A size effect is unearthed where both the frequency and severity of breaches scale with organisation size like $s^{0.6}$. Projections indicate that the total amount of breached information is expected to double from two to four billion items within the next five years, eclipsing the population of users of the Internet. This massive and uncontrolled dissemination of personal identities raises fundamental concerns about privacy.

\end{abstract}

\flushbottom
\maketitle
\thispagestyle{empty}

\section*{Introduction}

Both the Earth and humanity are often hit by extreme disasters characterised by their high severity and unpredictability \cite{pisarenko2010heavy}. Natural disasters -- such as floods, earthquakes, and hurricanes -- and man-made disasters -- such as financial crashes  \cite{Kindleberger00,Sornettecrash03}, nuclear power plant meltdowns \cite{WheatleyNuclear2015,sornette2013exploring}, military nuclear accidents \cite{Schlosser14}, the 2003 space shuttle explosion \cite{Leveson08} and many other extreme
industrial disasters \cite{Perron99,ChernovSor15} -- belong to this class. A recent entrant into this class are cyber attacks \cite{cyberattacks}, in which intellectual property can be exfiltrated, and operations of information systems -- or even critical physical infrastructures \cite{Coughlin11,Sanger12} -- disrupted. These cyber attacks can be perpetrated by cyber criminals, or even by state actors as acts of espionage or cyber warfare. Here we focus on personal data breaches, as a subset of cyber risks, where large amounts of personal information (i.e., name, social security number, address, email, date of birth, credit card numbers, usernames and passwords, etc.) are exfiltrated from organisations, typically for use in identity fraud. 

While appearing minor in view of other cyber risks, personal data breaches heavily impact both consumers and organisations \cite{harrell2013victims,ponemon2014}. For instance, the average financial loss due to the theft of a single piece of private data is estimated to be 213 USD \cite{ponemon2014}. Thus it is not surprising that data breaches can result in immediate negative impacts in the stock value of traded companies \cite{campbell2003economic, garg2003quantifying,acquisti2006there,gatzlaff2010effect}. Further, it is estimated that data breaches (including intellectual property) cost large companies more than 1 trillion US Dollars in 2008 \cite{McAfeeUnsecured}. In fact, the potential for major damage is so severe that the statistical properties of data breaches have been found to be similar to those of the most extreme disasters mentioned above \cite{maillart2010}. Acknowledging this, both governments, insurance companies, and organisations have started ranking cyber risk as one of the largest risks that they face (e.g., \cite{TelegraphLloyds,pwcCEO,WEF,UKgov,Allianz}).

Understanding the risk of disasters is essential for the proper design of infrastructures, emergency response planning, and for the construction of sound insurance policies \cite{Embrechts,embrechts1999extreme}. Modern approaches to this involve the systematic collection of high quality data and subsequent statistical risk analysis \cite{Embrechts,pisarenko2010heavy}. 

Early work on personal data breach risks \cite{maillart2010} has demonstrated the Pareto (Power Law) distribution function (df) of large breaches, which, having parameter $\alpha \approx 0.7$, is so \emph{extremely heavy tailed} that the largest observation is expected to be $(1-\alpha)/\alpha \approx 0.43$ times as large as the sum of the rest of the data \cite{Sornette}. For this reason, the mean (and higher moments) are infinite, and thus data breaches constitute an extreme statistical risk. 
Motivated by the major implications of such a dire characterisation, the fact that, in the past six years since \cite{maillart2010} was done, much more data has become available, and given the dynamic nature of the cyber space, we both significantly update and extend this analysis. With a current and much larger dataset, we confirm the heavier tailed breach df, and stable frequency since 2007. Going beyond this, we find that the breach df is in fact even heavier tailed than expected, and has a finite maximum. Further, we explicitly evaluate the connection between organisation size and breach risk.  

Specifically, we characterise the risk of large data breaches of personal data occurring at organisations. Damage is measured  by the number of individual information items (ids) exfiltrated. The recording of such data started in the 2000s, by an early online community, scouring media and other online reports \cite{datalossdb}, and has since become more mature with a variety of communities and organisations taking on the job \cite{ClearingHouse,Verizon}. Within the United States, this task has been aided by Freedom of Information (FOI) requests, and the onset of legal disclosure obligations faced by organisations having suffered a data breach (Data Breach Notification Act). However, perhaps due to the relatively short history of cyber risks, these databases are not without their weaknesses, including incompleteness, lack of standardisation, in-availability, and so on. Here, we have joined together the two largest databases \cite{datalossdb,ClearingHouse}, and filtered for events that occurred at an organisation (both public and private, businesses, universities, hospitals, etc.), within any country. This yielded 6,422 data breach events, each having in excess of 10 ids breached, between January 1st, 2000 and April 16, 2015, providing a solid basis for statistical analysis that allows us to estimate both the frequency and severity of events. For this modeling, we focus exclusively on large breaches (having in excess of fifty thousand ids). The severity is represented by a Pareto df with extensions allowing for the statistical hypothesis testing of (i) if there is a maximum breach size, (ii) if this maximum is growing, and (iii) if breaches tend to be getting larger over time. These models are tested against one-another and individually verified by rigorous diagnostics. Acknowledging the fact that breached data accumulates in the hands of cyber criminals, we also model the cumulative sum of large breaches over time. This brings together models for both frequency and severity, and provides a gloomy forecast for risk of data breaches, and thus the state of privacy, in the following five years.

Further, extreme statistical properties of disasters may often be related to a complex hierarchical underlying structure, in which cascades of failures develop into a broad df of sizes.
For instance, the Gutenberg-Richter (heavy-tailed power law) frequency-magnitude law for earthquakes
is thought to originate from the hierarchy of fault scales forming complex fault networks \cite{Scholz02}.
The proximate trigger of the 2008 financial meltdown can be attributed to a collapse of the hierarchical inter-bank network \cite{Soramakietal07}, when overnight loans backed by financial asset collaterals froze \cite{KacperczykSch10,Sieczkaetal11}. Financial bubbles develop from the interaction of many different agents at different scales-- computers, individuals, investment floors, and firms to currency blocks -- at different time scales \cite{SorHoh98}. 

Similarly, the Internet exhibits a socio-technological complexity that spans all levels of interactions. With fast-evolving hardware and software structures, and coupled with heterogenous and simultaneous interactions of millions of users, there are many potential points of failure. More specifically, data breach risk is related to underlying factors such as the \emph{attack surface}, which is the number of points where an attacker can extract data, as well as the volume and value of information assets that an organisation is guarding. As a single proxy risk factor for these variables, we consider organisation size, as defined by market capitalisation. With this measure, we unearth how both the frequency and magnitude of breach risk scale with organisation size -- thus quantifying the effective \emph{risk surface} of an organisation.
 
\section*{Data, Results \& Methods}

The risk of data breaches is considered where each breach has an event time, being the date at which it was reported (to a governmental body or media outlet), and a size, being the number of pieces of personal information breached. 

The risk is decomposed into frequency and severity components, and these two components are studied separately. Only events above a threshold size are considered. This is sensible for (at least) two reasons: Firstly, large breaches dominate the total number of breached ids. For instance, although events with size above $5\times10^4$ only represent less than 10 percent of events, the total number of ids' breached across these large events is above 99 percent of the total across all events (Tab.~\ref{tab:rate}). Thus, we want good data for reliable statistics about large breaches. This relates to the second reason: large breach events are more visible, and thus the data are more complete and reliable in this range. Including smaller events in the data set worsens data quality and may bias statistics. We choose the breach size threshold to be $5 \times 10^4$, and breaches above this threshold will be referred to as \emph{large breaches}. To further select reliable data, we study events occuring from January $1^{st}$, 2007 until April $15^{th}$, 2015, being the most recent date available at the time of analysis. Data is available prior to 2007, however the number of events are relatively few, and the statistics less consistent. 

It is important to note that almost 40 percent of events have an unknown size (Tab.~\ref{tab:rate}). If we exclude these events, and if these events with unknown size tend to be small (i.e., falling beneath the threshold of $5\times10^4$), then there is no problem. This may well be the case. If they also tend to be large, then we underestimate the frequency. If they tend to be large, and follow a different distribution than the rest of the sample, then this will bias the estimate of the distribution. However, given a lack of covariates to identify if there is a way to predict if an event will have an unknown size, we can neither evaluate the effect of this, neither perform any meaningful imputation  (i.e., estimate the missing values in a way that is consistent with the result of the data). That is, any imputation will only sample these events from the distribution identified by the observed events, and thus have no impact on the distribution. Further, this will require an assumption about what proportion of the events with unknown size will be above the threshold. Thus, we simply omit the events with unknown size, providing an optimistic quantification of the risk. 

To enable rigorous statistical modeling, we introduce some notation. We consider the \emph{large breach sizes} $\{ x_i,~i=1,2,\dots,n=619 \}$, having limited size $5\times 10^4 < x_i \leq \nu$, with given lower threshold $5\times 10^4$, and parameter $\nu$ gives the unknown maximum breach size. The data $x_i$ are ordered by their event time $0<t_{1}< t_{2}<\dots<t_{n}$, where the clock starts at $t=0$, being January $1^{st}$, 2007, and one unit of $t$ is a year.

\begin{table}[!h] 
	\begin{center}
	\scalebox{1}{
	\begin{tabular}{c | c c c c c}
	\toprule
	  Category		& n 	& Total Breach		& Monthly Rate	& 	Annual Rate	 &	GLM			\\
	 \midrule

	  US 			& 6142	& $1.189\times10^9$	&$62.6~(13.1)$	&   	$751.0~(110.9)$	 & 		-		\\
	  US$>5\times10^4$	& 407	& $1.174\times10^9$	&$4.25~(1.82)$	&	$49.5~(6)$	 & $4.59~(0.5)\text{;}~-0.08~(0.1), 0.43$\\
	  Non-US 		& 1978	& $0.794\times10^9$	&$24.6~(18.0)$	&	$296.0~(199.2)$	&		-		\\
	  Non-US$>5\times10^4$  & 186	& $0.788\times10^9$	&$2.48~(1.65)$	&	$26.0~(11.4)$	& $1.64~(0.4)\text{;}~0.19~(0.1), 0.02$\\
	  \midrule

	  All			& 8574	& $1.983\times10^9$	&$87.2~(27.1)$	&	$1046.5~(292.1)$&		-			\\
	  All$>5\times10^4$	& 619	& $1.962\times10^9$	&$6.29~(2.65)$	&	$75.5~(10.38)$& $5.44~(0.6)\text{;}~0.18~(0.12), 0.13$\\
         \bottomrule
	\end{tabular}	
	}
	\end{center}
	\caption{ The number of events (n), total number of breached ids (Total Breach), average monthly count (Monthly Rate) and standard deviation of monthly counts, average annual count (Annual Rate) and standard deviation of annual counts, and GLM summary. The generalised linear model (GLM) summary provides the intercept parameter (events per month) with standard error, and GLM slope parameter (events per year) with standard error and p-value. These statistics are given for events occurring to US firms (US), to non-US firms, and to all firms together (All). Statistics were taken on the window of January $1^{st}$, 2007 until April $15^{th}$, 2015. For each of the aforementioned categories, the statistics are given for all sizes (including 2683 events with unknown size within the US, and 526 events with unknown size outside of the US), and for events with breach in excess of fifty thousand ids. ($>5\times10^4$). }
	\label{tab:rate}
\end{table}

\subsection*{Data Breach Frequency}

The rate (frequency) of breach events is studied, with relevant statistics presented in Tab.~\ref{tab:rate}. According to a linear regression of monthly counts over time (Poisson Generalised Linear Model with identity link \cite{GLM},) the rate of large events has been stable within the US, and growing significantly outside of the US -- driving almost significant ($p=0.13$) growth when all countries are taken together. However, this growth is $0.18$ events per year, which is only a fifth of a percent of the total annual rate, thus being practically insignificant. This apparent stability runs counter to the view that cyber risks are worsening. Next, we consider the dynamics in the size of large breaches, which provides a less reassuring message.

\subsection*{Data Breach Severity}

\begin{figure}[h!] 
\begin{center}
\centerline{\includegraphics[width=16cm]{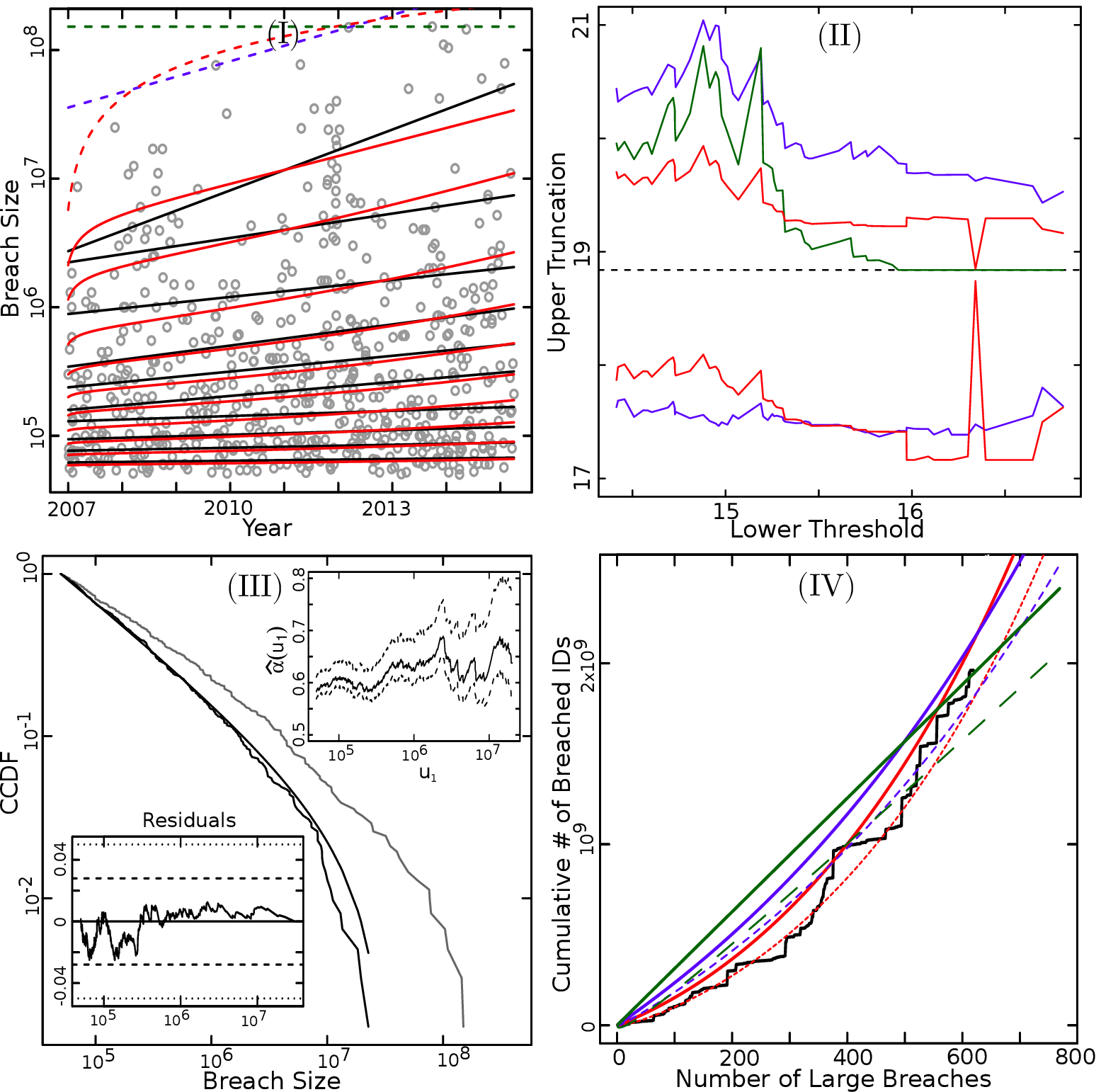}}
\caption{ Panel (I) plots large events (above $5\times10^4$) over time from January $1^{st}$, 2007 until April $15^{th}$, 2015, as well as the $p=(0.1,0.2,\dots 0.9)$ and $0.95$ quantile levels of a linear quantile regression (black), and $D_2$ (eq.~\ref{eq:DTP}; red), estimated on the data. The upper endpoints of $M_1$ (green), $M_2$ (blue) and $M_3$ (red) are given by the dashed lines. Panel (II) plots EVT estimates of the maximum of the natural log of the data. $M_1$ is given in green. $M_2$ (blue) and $M_3$ (red), having growing maximum, have the maximum plotted at both January $1^{st}$, 2007 (lower) and at April $15^{th}$, 2015 (upper). Panel (III), in the main frame, plots the empirical complementary cumulative distribution function (CCDF; rough black line) of the transformed breach sizes (for the second alternative model), the DTP CCDF (smooth black line) $\widehat{\alpha}=0.56$) estimated on the transformed breach sizes, and the empirical CCDF for the untransformed data (grey line). The inner left frame plots the ``residual'' distances between the empirical and estimated CDFs (the two black lines), with $(0.1,0.25,0.5,0.75,0.9)$ quantiles plotted for the Kolmogorov Smirnov df. The inner right frame is a sequence of DTP shape parameter estimates, and standard errors, on the transformed data, for increasing lower truncations, $u$. Panel (IV) plots the observed cumulative sum of the 619 large breaches occurring from January $1^{st}$, 2007 until April $15^{th}$, 2015 (rough black). The horizontal axis extends to 800, which is the number of events to be expected about 2.5 years into the future. The green, blue, and red lines provide the expected cumulative sum under the estimated null, first, and second alternative models, respectively. The lower standard error for each of the three curves is plotted in the same color, with the null, first, and second alternative models having long dashed, medium dashed, and densely dashed lines respectively. The upper standard error will be the same distance from the expected cumulative sum, but on the positive side.}
\label{fig:F}
\end{center}
\end{figure}

The dynamics of the df of large breach sizes over time are studied. Given the growing amount of data being stored online, and the evolution of cyber crime methods, the severity of breaches is expected to increase. As an initial diagnostic, the observed cumulative sum of large breaches over time is shown, in Fig.~\ref{fig:F} (panel IV), to curve up -- indicating growth in the mean breach size. We thus consider three possibilities: (i) there is a maximum possible breach size, (ii) this maximum is growing, and (iii) large breaches are becoming larger over time. 

\subsubsection*{Detecting The Maximum Breach Size}

We use Extreme Value Theory (EVT) \cite{Embrechts} to determine if a maximum breach size exists, and if so, how it has evolved. EVT provides a standard framework for such statistical inference ``beyond the largest observation''. We exploit a general EVT theorem that roughly states: for large values of random variable $X$ being in excess of a sufficiently large truncation $u$, the \emph{Generalized Pareto Distribution} (GPD) approximately models the tail of the df of $X$,
\begin{equation}
 Pr\{X-u \leq x |t,~X-u>0 \} \approx 1-\left(1-\xi x/\beta(t) \right)^{-1/\xi} ~,~\text{with}~~\beta(t)>0~~ ,~ \text{for ~$\xi \neq 0$}~, 
  \label{eq:GPD}
\end{equation}
with $0 \leq x$ without upper bound if the \emph{Extreme Value Index} $\xi $ is positive and $0\leq x \leq -\beta(t)/\xi$ for $\xi <0$. In this later case,  a finite maximum exists for $X$,
\begin{equation}
\nu(t) = u-\beta(t)/\xi <~~~ \infty~,
\label{eq:upperEndpoint}
\end{equation}
which may vary with time when the scale parameter $\beta$ is time dependent. 

To characterize the maximum breach of the (natural) log of breach sizes, $\nu(t)$, we consider the following statistical hypotheses and their corresponding parameterisations of~\eqref{eq:GPD},
{\setlength\arraycolsep{0.1em}
\begin{eqnarray}
 M_0&:&~\text{no maximum size is detected}~(\xi > 0)~\text{;} \label{eq:M0} \\
 M_1&:&~\text{there is a constant log-maximum}~(\xi < 0,~\beta(t)=\beta_0)~\text{;} \label{eq:M1} \\
 M_2&:&~\text{the log-max grows linearly in time}~(\xi < 0,~\beta(t)=\beta_0+\beta_1 t,~\beta_1>0)~\text{;} \label{eq:M2} \\
 M_3&:&~\text{the log-max grows sub-linearly in time}~(\xi < 0,~\beta(t)=\beta_0 + \beta_1 \text{ln}(t),~\beta_1>0)~\text{.} \label{eq:M3} 
\end{eqnarray}
}

For the GPD approximation~\eqref{eq:GPD} to be good, one wants $u$ to be as large as possible, but at the same time one wishes to have a large sample. Thus, we take estimates at the lowest value of $u$ at which parameter estimates stabilize. The GPD~\eqref{eq:GPD} with its parameterisations~(eqs.~\ref{eq:M1}-\ref{eq:M3}) were estimated on the natural log of the breach sizes (\emph{Peaks Over Threshold (POT) Estimation}~\cite{Coles}), for lower thresholds ranging from $u=14.4$ (having 102 points above) to $u=16.8$ (having 20 points above).  Considering that the estimated maximum is approximately stable for $u>15.5$ (Fig.~\ref{fig:F}, panel (II)), estimates are taken at $u=15.5$ (having 50 points above), and recorded in Tab.~\ref{tab:EVT}. 

The main insights about the behaviour of the maximum size gained from this estimation are visualized in Fig.~\ref{fig:F} (panels (I) and (II)).  For a range of lower thresholds $u$, the estimated maximum breach size are ``hugging'' the data, implying the existence of a highly significant upper truncation which the data has already reached. Moreover, we find a highly significant upward growth of this maximum size. In further detail, for $M_1$, the estimated shape parameter $\xi$ is already significantly negative ($\widehat{\xi}=-0.36~(0.1),~p\approx0.001$) with small $u=14.4$, and achieves values below $-1$ for $u>16$, indicating a highly significant maximum. Both $M_2$ and $M_3$ exhibit significant growth in the maximum over time ($p<0.001$ in Tab.~\ref{tab:EVT}), and have superior likelihood to $M_1$ by the likelihood ratio test (LRT), having $p=0.08$ and $p=0.05$, respectively. Finally, $M_3$, has superior log-likelihood to $M_2$ for all $u>15.5$ with the same number of parameters. Thus, the best model for the growth of the maximum (obtained by exponentiation of the log-maximum) is the sublinear power,
\begin{equation}
\exp(\nu(t))\propto t^{-\beta_1/\xi}\approx t^{0.83}.
\end{equation}
Indeed it may seem obvious that here -- as in any natural (finite) system -- there is a maximum size. Further, given the flow of users and data online, and the growth of giant IT companies, it is sensible that this maximum possible breach size is increasing. However, to quantify this is of high importance to policymakers and (re-)insurance firms who care about the aggregate risk, which is dominated by the largest observations of such heavy tailed df. It also makes a major theoretical distinction as for this model with a finite maximum all moments are finite, whereas with an infinite maximum all moments are infinite.

\begin{table}[!h]
	\caption{EVT peak-over-threshold (POT) estimates of the three models (eqs.~\ref{eq:M1}-\ref{eq:M3}) are presented for lower threshold $u$, with loglikelihood (ll), and estimated shape $\xi$ (with standard error), scale intercept $\beta_0$ (with standard error) and scale slope $\beta_1$ (with standard error and p-value).}
	\begin{center}
	\scalebox{1}{
	\begin{tabular}{r | c c c c c c c c }
	\toprule
     Model	&u	& ll    &$\xi$& $\beta_0$ &  $\beta_1$ 		  \\ 
\midrule
$M_1$		& $15.5$&$-60$&$-0.61~(0.18)$&$2.24~(0.47)$&     $=0$     \\
$M_2$		& $15.5$&$-57$&$-0.65~(0.15)$&$1.35~(0.39)$&$0.18~(0.06),~0.001$ \\
$M_3$		& $15.5$&$-56$&$-0.78~(0.15)$&$1.60~(0.26)$&$0.65~(0.15),~0.7\times 10^{-6}$ \\
\bottomrule
	\end{tabular}	
	}
	\end{center}
	\label{tab:EVT}
\end{table}

\subsubsection*{The Distribution of Large Breach Sizes}

The df of large breach sizes is estimated to quantify the severity of data breach risks and their dynamics. We model the large breach sizes by a doubly truncated Pareto (DTP) df typical for modeling extreme risks,
\begin{equation}
 F_{DTP}(x|t)= \frac{1-(x/\widetilde{u})^{-\alpha(t)}}{1-(\widetilde{\nu}/\widetilde{u})^{-\alpha(t)}}~,~~0< \widetilde{u} < x\leq \widetilde{\nu}~,~\alpha(t)>0~,
 \label{eq:DTP}
\end{equation}
having \emph{shape parameter} $\alpha(t)$ potentially varying in time. Rather than working directly with~\eqref{eq:DTP}, for convenience we work with the natural logarithm of the data, $Y=$ln$(X)$, which follows a doubly truncated Exponential (DTE) df,
\begin{equation}
 F_{DTE}(y|t)= \frac{1-\text{exp}(-\alpha(t) (y-u) )}{1-\text{exp}(-\alpha(t) (\nu-u) )}~,~~0< u < y\leq \nu~,~\alpha(t)>0~,
 \label{eq:DTE}
\end{equation}
with $u=$ln$(\widetilde{u})$ and $\nu=$ln$(\widetilde{\nu})$.

We have already determined that a significant and growing maximum breach size exists. However, it is not yet known what other trends have been present within breach severity. We thus consider statistical hypotheses about trends in the df of large breaches, and their corresponding parameterisations in~\eqref{eq:DTE}:
{\setlength\arraycolsep{0.1em}
\begin{eqnarray}
  D_0&:&~\text{the df has a fixed max ($\nu(t)=\nu_{0}$) and is stationary ($\alpha(t)=\alpha_{0}$) ;} \label{eq:D0}\\
  D_1&:&~\text{the df has a fixed max ($\nu(t)=\nu_{0}$) and becomes more heavy tailed ($\alpha(t)=\alpha_{0}+\alpha_{1} t~,~\alpha_1<0$) ;}  \label{eq:D1} \\
  D_2&:&~\text{the maximum log-breach grows sub-linearly ($\nu(t)=\nu_{0}+\nu_{1}\text{ln}(t)~,~\nu_1>0$),   } \nonumber \\   
  &&\text{  and the df becomes more heavy tailed ($\alpha(t)=\alpha_{0}+\alpha_{1} t~,~\alpha_1<0$) .} \label{eq:D2}
\end{eqnarray}
}
The hypothesis $D_2$~\eqref{eq:D2} contains \eqref{eq:M3} where $\nu_0=u-\beta_0/\xi$ and $\nu_1=-\beta_1/\xi$. The hypotheses $D_0$~\eqref{eq:D0} and $D_1$~\eqref{eq:D1} overlap with testing of the previous hypothesis tests~(\ref{eq:M0}-\ref{eq:M3}), providing an opportunity for confirmation of results about the maximum with the specific DTE model~\eqref{eq:DTE}.


The DTE~\eqref{eq:DTE} with specifications (\ref{eq:D0}-\ref{eq:D2}) were estimated by Maximum Likelihood (ML). For $D_2$, the maximum was given by $\nu_0\approx13.63$ and $\nu_1\approx 0.84$, computed from the EVT estimates ($M_3$ in Table~\ref{tab:F_params}).

The estimation (Table~\ref{tab:F_params}) finds that large breaches have grown larger (the tail has become heavier), and confirms that a maximum breach size exists, and is increasing. More specifically, it is confirmed that a maximal breach size is clearly present by comparing the likelihood of $D_0$ with and without finite maximum ($p\approx10^{-7}$ by the LRT). It is also found that large breaches have become heavier tailed, with the shape parameter of $D_1$ significantly decreasing ($p=0.04$) at a rate of $-0.027$ per year. This extension of the model significantly increases the quality of fit ($D_1$ has superior likelihood to $D_0$ with $p_{LRT}=0.006$). Finally, in confirmation with the EVT results, the model \eqref{eq:D2} is found to be best, where the growing maximum further improves the result of $D_1$ ($D_2$ has superior likelihood to $D_1$ with $p_{LRT}=0.007$). 

 These results are all quite striking. For instance, all estimates, having shape parameter $\alpha<1$, are so extremely heavy tailed that, without a finite maximum, the mean of this model would be infinite. For $D_0$, having shape parameter $\alpha=0.47$ and maximal breach $1.52\times10^8$, the expected large breach size is $3.1\times10^6$, with even larger standard deviation, $1.3\times10^7$~\eqref{eq:DTPMoments}. For $D_2$ in 2015 ($t=8$), having $\alpha(8)=0.364$ and maximum exp$(\nu(t))=2.24\times10^8$, the expected large breach size becomes twice as large as for $D_0$. Further, under $D_2$, given a breach size is in excess of fifty thousand, there is a 10 percent chance that it exceeds ten million ids.
 
 In Fig.~\ref{fig:F} (panel III), diagnostics are given to demonstrate the validity of $D_2$. For this, the data are transformed to be stationary and then standard diagnostics are performed. In detail, if $D_2$ is true, then the log breach random variable $Y=\text{ln}(X)$ is from model~\eqref{eq:DTE} with non-stationary parameters~\eqref{eq:D2} with estimated values in Tab.~\ref{tab:F_params}. Thus, the transformed data,
\begin{equation}
 \widetilde{Y_i}=Y\times\frac{\alpha(t)}{\alpha_{0}}~\rvert~t\sim \text{DTE}\left( \alpha_{0}~,~\nu^*(t) \right)~ (\ref{eq:DTE}),~~~~
 \nu^*(t)=\frac{\alpha(t)}{\alpha_{0}}\times\nu(t)\approx\nu_0~,
 \label{eq:transform}
\end{equation}
 will be approximately stationary with df equal to that of the log breaches $Y$ at time 0 (January $1^{st}$, 2007). Estimating the model~\eqref{eq:DTE}, with the $D_0$ specification, on the transformed data~\eqref{eq:transform}, yields an estimate of $\widehat{\alpha}=0.58$, which agrees with the estimate $D_2$ at $t=0$ ($\alpha(0)=\alpha_0=0.57$), indicating that the transformation is valid. Further, in Fig.~\ref{fig:F} (panel III), the empirical complementary CDF of the transformed data are found to be well described by this estimated stationary model, as evidenced by small residuals between the empirical and estimated complementary CDFs -- the Kolmogorov-Smirnov \cite{KSTest} has p-value of 0.78 -- as well as the consistency of the shape parameter $\alpha$ for lower truncations $u$ ranging from $5\times10^4$ to $5\times 10^7$. 

 The above results are strengthened and confirmed by obtaining similar results with a more flexible method. We used quantile regression \cite{Koenker} which, rather than specifying a parametric model, independently fits linear regressions to each quantile. In support of the dynamic specifications $D_1$ and $D_2$, Fig.~\ref{fig:F} (panel I) shows good agreement between the linear quantile regressions and the growing quantiles of $D_2$. Further, the slope, standard error, and p-value of the
 linear regressions of the 0.5 and 0.9 quantiles are $0.083~(0.038),~0.03$ and $0.145~(0.07),~0.04$, respectively. Thus the quantile regressions are significantly increasing, providing strong additional evidence that large breaches are getting larger.
 
  \begin{table}[!h]
	\caption{ Parameter estimates, standard errors (Monte Carlo with 1000 repetitions), and p-values for the significance of slope parameters (Monte Carlo with 1000 repetitions), of model \eqref{eq:DTE} with parameter specifications eqs.~\ref{eq:D0}-\ref{eq:D2}. The $D_0$* model has no finite maximum.	}
	\begin{center}
	\scalebox{1}{
	\begin{tabular}{r | c c c c c c c }
	\toprule
     Parameter	 	& ll	&  $\alpha_0$     &  $\alpha_1$ 			&  $\nu_0$	&$\nu_1$	\\ 
\midrule
$D_0$		&$-1020.7$&$0.47~(0.017)$ &	$=0$			& $18.839~(0.2)$& $=0$			\\
$D_0$*			&$-1032.9$&$0.51~(0.02)$  &	$=0$			& $=\infty$	& $=0$			\\
$D_1$		&$-1017.0$&$0.58~(0.05)$  &$-0.027~(0.01),~0.002$	& $18.839~(0.2)$& $=0$  		\\
$D_2$		&$-1013.3$&$0.57~(0.05)$ &$-0.025~(0.01),~0.014$	& $=13.63$ 	&$=0.84$   		\\
\bottomrule
	\end{tabular}	
	}
	\end{center}
	\label{tab:F_params}
\end{table}

 \subsection*{Cumulative Risk \& Future Projections}
 
 Due to the large-scale sharing of breached data, e.g., by sophisticated underground markets \cite{franklin2007,Rand}, breached personal information is concentrated, enabling efficient subsequent identity fraud \cite{harrell2013victims}. Thus, privacy {\it erodes} with the growth of the cumulative number of breached ids. For this reason, to understand the long term risk of data breaches to privacy, it is crucial to study the cumulative amount of breached information. The cumulative sum measure brings together both frequency and severity in a convenient way for the study of past and future evolution of risk.
 
 In Fig.~\ref{fig:F} (panel IV), the observed cumulative sum, $C_n=\sum_{i=1}^n x_i$, of the $n=619$ large breaches occurring from January $1^{st}$, 2007 until April $15^{th}$, 2015, is plotted. If both the statistics of frequency and severity were stable over time, the cumulative sum would grow linearly in $n$. That the observed cumulative sum curves upward indicates a growing mean, as featured in the $D_1$ and $D_2$ specifications of~\eqref{eq:DTE}. To compare the observed data with the models, the expected cumulative sum $E[C_n]=nE[X]$, and its standard devation, are plotted for the estimated models. This simply requires computing the first two moments of the DTP (\ref{eq:DTP}):
  \begin{equation}
  \text{E}[X]= \frac { \alpha }{\alpha-1 }\left[ \frac{ u^{1-\alpha}-\nu^{1-\alpha} }{ u^{-\alpha}-\nu^{-\alpha} } \right]~,
  ~~~~\text{E}[X^2]= \frac { \alpha }{\alpha-2 }\left[ \frac{ u^{2-\alpha}-\nu^{2-\alpha} }{ u^{-\alpha}-\nu^{-\alpha} } \right]~.
  \label{eq:DTPMoments} 
  \end{equation}
  
However, there is an important subtlety in comparing the observed and expected curves. The relevance of the expected sum to the observed sum relies upon the Central Limit Theorem (CLT). For infinitely large $\nu$, the mean is infinite and thus the CLT never converges. In this case, the cumulative sum would grow superlinearly, $C_n\sim u ~n^{1/\alpha}$, rather than linearly \cite{Sornette}. For instance, for $\alpha=0.5$, this curve would have exceeded the upper boundary of the Fig.~\ref{fig:F} (panel IV) within the first $n=200$ observations. A rule of thumb for when the CLT holds, and the cumulative sum grows linearly in $n$ rather than superlinearly, is for samples larger than $n^{*}\approx (\frac{\nu}{u})^\alpha$ \cite{Sornette}. Here where $u=5\times10^4,~\nu=1.6\times10^8$ and $\alpha\approx0.5$, the crossover point is $n^{*}\approx50$. 
This means that the observed upward bend in the cumulative sum occuring after $n^{*}$ is not due
to the superlinear growth resulting from the heavy-tailed CDF but rather results from 
the non-stationarity of the process. Comparing the observed and expected curves: $D_0$, which grows linearly, fails to capture the trend; $D_1$, which is convex due to an increasingly heavy tail, partially captures the trend; and $D_2$ is again the best model, capturing the trend well due to both an increasing maximum and increasingly heavy tail.
  
For future projections, we make use of all fitted models. The previous analysis was conditional upon knowing the number of breaches. To more properly quantify the uncertainty of projections, the annual number of breaches $N_t$ is treated as random. The annual sum, 
\begin{equation}
Y_t=X_1+\dots + X_{N_t}~,
\label{jythnb}
\end{equation}
(with all $X_i$ and $N_t$ independent) is called a \emph{Compound Process}~\cite{Mikosch,Wuthrich}. The mean and variance are given by,
    \begin{equation}
 \text{E}\left[ Y_t \right]=\text{E}\left[X_t\right]\text{E}\left[N_t\right]~~,~~~~\text{Var}(Y_t)=\text{E}\left[N_t\right]\text{Var}(X_t)+\text{E}\left[X_t\right]^2\text{Var}(N_t)~~,
  \label{eq:CompoundMoments} 
  \end{equation}
  which are computed for the coming five years of 2015-2019 in Tab.~\ref{tab:projections}, where the four months of data after January $1^{st}$, 2015 were excluded for convenience. One observes that the cumulative breach $C_t$, currently at a level of around $1.816\times 10^9,$ is expected to more than double in the next five years under the nonstationary $D_1$ and $D_2$ model specifications.  
 
   \begin{table}[!h]
	\begin{center}
	\scalebox{1}{
	\begin{tabular}{r c c | c c c c c }
	\toprule
Model 	 &Quantity		&	$2014$	&   $2015$   &  $2016$		&  $2017$	&$2018$ 	& $2019$		\\ 
\midrule	
$D_0$	&$Y_t\times10^{-8}$	&	-	& $2.37~(1.14)$ & $2.37~(1.14)$ & $2.37~(1.14)$ & $2.37~(1.14)$ & $2.37~(1.14)$ 	\\
$D_0$	&$C_t\times10^{-8}$	&	$=18.16$& $20.5~(1.14)$ & $22.9~(1.61)$ & $25.3~(1.94)$ & $27.6~(2.28)$ & $30.0~(2.55)$ 	\\
$D_1$	&$Y_t\times10^{-8}$	&	-	& $3.73~(1.52)$ & $4.18~(1.63)$ & $4.67~(1.74)$ & $5.21~(1.86)$ & $5.81~(1.99)$ 	\\
$D_1$	&$C_t\times10^{-8}$	&	$=18.16$& $21.9~(1.52)$ & $26.1~(2.22)$ & $30.7~(2.82)$ & $36.0~(3.38)$ & $41.8~(3.92)$ 	\\
$D_2$	&$Y_t\times10^{-8}$	&	-	& $4.96~(2.12)$ & $5.97~(2.48)$ & $7.16~(2.87)$ & $8.56~(3.31)$ & $10.2~(3.79)$ 	\\
$D_2$	&$C_t\times10^{-8}$	&	$=18.16$& $23.1~(2.12)$ & $29.1~(3.27)$ & $36.3~(4.35)$ & $44.8~(5.47)$ & $55.0~(6.65)$ 	\\

\bottomrule
	\end{tabular}	
	}
	\end{center}
	\caption{Future expected annual breaches (and standard deviation)~\eqref{eq:CompoundMoments} 
of the annual sum $Y_t$ (\ref{jythnb}) and cumulative breach $C_t$ since 2007
are presented with annual rate estimates from Table~\ref{tab:rate}, and DTP df~\eqref{eq:DTP} with parameterisations given in eqs.~\ref{eq:D0}-\ref{eq:D2}, and parameter values in Table~\ref{tab:F_params}. All values in the table are divided by $10^8$. These estimates assume a constant rate of 75.6 large events per year with annual variance of 229 (computed across 2007 until 2015). The US rate estimate is 50.4, and thus US estimates may be obtained by scaling the numbers by $49.5/75.6\approx2/3$ (the US total is $11.89\times10^8$, which is also about $2/3$ of the total of $19.62\times10^8$). }
	\label{tab:projections}
\end{table}

\subsection*{Data Breach Risk \& Organisation Size}

  Larger organisations tend to have more attractive assets to motivate a cyber attack, present a larger \emph{attack surface} making penetration easier, and once penetrated, contain more data to be exfiltrated \cite{simon2013too}. Thus, larger organisations will be both more frequently and severely victimised than smaller ones. To quantify the relationship between risk and organisation size, we use market capitalisation (MCAP) as a proxy for size, and compare organisations of different size with past observed breaches. Specifically, we consider 4,950 firms publicly traded on the New York Stock Exchange (NYSE), American Stock Exchange (AMEX), and the Nasdaq~\cite{NASDAQCompanyList}. For victimised firms, MCAP was taken the day before the data breach incident, to avoid possible subsequent devaluations due to the breach (e.g, see \cite{campbell2003economic}), and to avoid price changes that often occur in the highly dynamic stock market. For non-victimised firms, the MCAP was taken at June $1^{st}$, 2014. All MCAP values were then inflation adjusted to 2014 US Dollars. Within the dataset studied here, there were 735 events across 400 of these companies, between 2006 and 2015. 
  
\begin{figure}[h!]
  \begin{center}
  \centerline{\includegraphics[width=18cm]{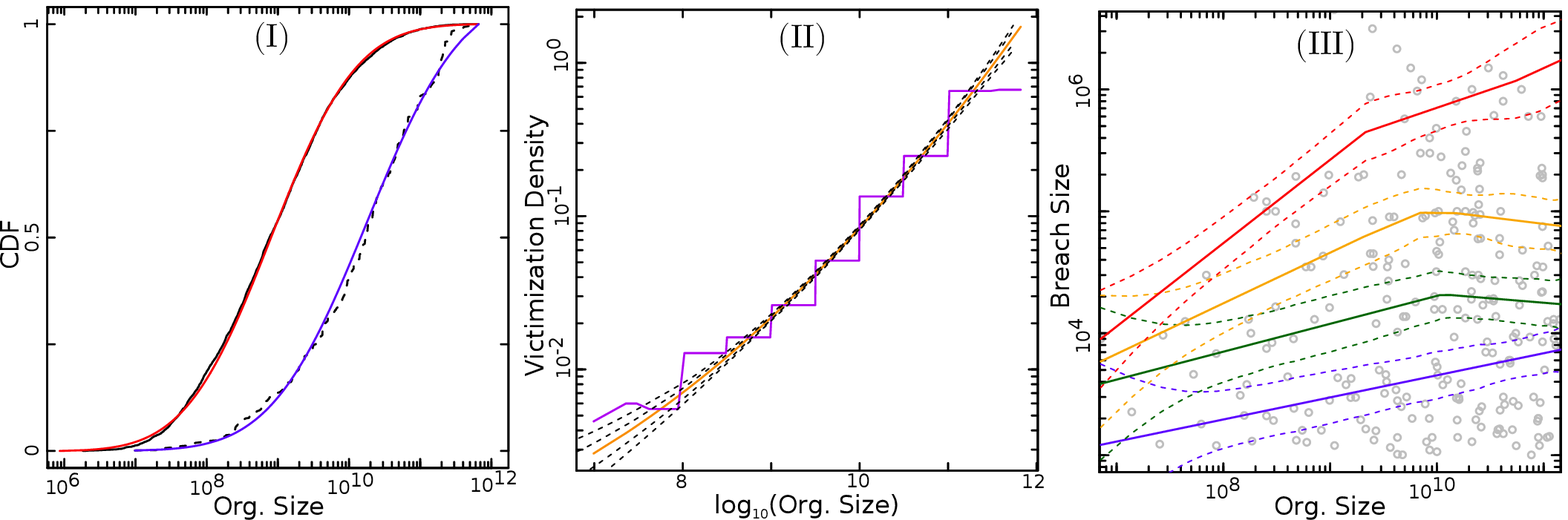}}
  \caption{The effect of firm size (market capitalisation) on the risk of breaches: Panel (I) provides the empirical (black) and estimated Lognormal CDFs for organisation sizes (red), and victimised organisation sizes (blue). The estimated parameter values and standard errors are $(\mu=20.3~(0.06), \sigma=2.1~(0.06))$ and $(\mu=23.6~(0.14), \sigma=2.5~(0.1))$, respectively. Panel (II) plots the estimated pdf defining the probability that firms of different log-size are victimised. The purple kinked line is computed by taking histogram estimates of the population and victim log-size densities and plugging them into (eq. \ref{eq:victimPDF}). The orange line does the same but with the Lognormal estimates. Monte Carlo $(0.1,0.25,0.5,0.75,0.9)$ quantiles (dashed black) of the estimated (orange) pdf are given by repeated subsampling from the estimated Lognormal CDFs and recomputation of~\eqref{eq:victimPDF}. Panel (III) plots the log of breach sizes in excess of $10^3$ ($n=298$), that occurred at US organisations versus log org. size. The lines are linear quantile regressions of this data, where change points are automatically detected (additive quantile regression) when apparent in the data. The $q=$ 0.3, 0.5, 0.7, and 0.9 quantile regressions are given. The estimated upward slopes, standard errors, and p-values, of these quantiles, for org. size below $10^{10}$, are: $q=0.3$: 0.66 (0.16) 0.00; $q=0.5$: 0.40 (0.16) 0.013; $q=0.7$: 0.23 (0.14) 0.02; $q=0.9$: 0.18 (0.12) 0.39.}
  \label{fig:Size}
  \end{center}
  \end{figure}

  To model the cumulative distribution function (CDF) of the size of all firms ($n=4,950$), and the size of firms given that they have been victimised ($n=735$), we consider a Lognormal df. This model choice has substantial justification, being a good model for a variety of size measures ranging from profits, to MCAP, to the number of employees \cite{Simon1958,Cabral2003,Gupta2007} and because it asymptotically encompasses the Zipf law family \cite{MalevPisSor11} often advocated based on the interplay between proportional growth and firm birth and death events \cite{Sornette1997,Sornette1998,Sornette1998_2,Saichev2009}. The df is truncated from above and below by the observed maximum and minimum of $10^6$ and $6.6\times10^{11}$, respectively. The estimated dfs are plotted with the empirical dfs in Fig.~\ref{fig:Size} (panel (I)), demonstrating reasonable goodness of fit. The CDF of victims, being shifted about 1 decade to the right of its unconditional counterpart, gives much higher probability to larger firms being victimised. This is further studied in the next subsection, and the relationship between firm size and breach size is presented afterwards.

  \subsubsection*{Breach Frequency \& Firm Size}
  
  The effect of firm size on the frequency of breach events is now studied. Despite the fact that only 10 percent of the publicly traded firms have been victimised, nine of the ten largest have been, and often multiple times (e.g., both Apple Inc. and IBM 6 times, Facebook 8 times, Bank of America 8 times, HSBC 17 times, General Electric 2 times, and Wal-Mart 6 times). Thus larger firms, despite being rare, are frequently victimised. The relative frequency at which firms with a given log(MCAP) are victimised is quantified by the victimisation pdf,
    \begin{equation}
  \text{Pr}\{\text{Firm log(MCAP) size$=x$}|\text{Firm is victimised}\} \propto \frac{f_{\text{Victim}}(x)}{f_{\text{Population}}(x)}~~,
   \label{eq:victimPDF}
  \end{equation}
  which is proportional to the ratio of the victim and population log size densities. Using the Lognormal estimates from the previous section as well as histogram density estimates, the victimisation pdf~\eqref{eq:victimPDF} is plotted in panel (II) of Fig.~\ref{fig:Size}. Both the histogram and parametric estimate suggest approximate linear growth, with slope 0.6, for sizes between $10^8$ and $10^{11}$. Thus the probability of victimisation grows with size as $\sim s^{0.6}$.
 
  The way that frequency (and breach size, to be seen below) scale with firm size can be interpreted as a quantification of an effective \emph{risk surface}. A 3D object with volume $c$ has surface area proportional to $c^{2/3}$. In d dimensions, the exponent is $(d-1)/d$. Thus the exponent 0.6 would correspond to a (fractal) dimension of $d_f=2.5$, implying that the risk surface scales with firm size in a way between that of the circle delineating a disk and the sphere bounding a ball.
 
  Another line of interpretation acknowledges that firms are composed of sub-units, and that breaches occur at this level. An example of this is Sony Corporation, which suffered a data breach on its PlayStation Network (Sony Computer Entertainment Division) in 2011, and a separate massive attack on Sony Picture Entertainment in late 2014. In \cite{Amaral1998}, it was found that the number of subunits in a firm scales with firm size as $\sim s^{1/3}$ and that the size of the largest sub-unit within a firm scales like $\sim s^{2/3}$. A pleasant, but simplistic, connection between this scaling and our result would be that the probability of attacks is proportional to the size of the largest unit, which is arguably the most visible and vulnerable. In reality, the number of subunits may also play a role.
  
  \subsubsection*{Breach Size \& Firm Size}
    
  We now quantify the way in which breach size scales with firm size. Companies with larger market capitalisation, tending to have more customers, retain more personal data that can be leaked. Further, personal data are increasingly considered as the {\it ore} that companies {\it mine} to extract consumer profiles and enhance online commerce \cite{simon2013too}. Thus, the value (and size) of personal data assets are increasingly likely to be reflected in market capitalisation~\cite{cauwels2012quis}. 
  
  In Fig.~\ref{fig:Size} (panel III), the $298$ breaches with size in excess of $10^3$, that occurred at publicly traded US firms, are plotted versus organisation size (MCAP). Further, linear quantile regressions of this data, with automatically detected changepoints (additive quantile regression~\cite{Koenker}), are plotted. One sees that the largest breaches occur at larger organisations. That is, the quantile regressions for quantiles 0.5 and above significantly increase ($p<0.05$) with firm size, until this size effect saturates for firm sizes above $10^{10}$. In particular, for the 0.9 quantile, breach size increases with slope 0.66, indicating that the largest breach sizes scale with firm size as$\sim s^{0.66}$. 
  
  It is also apparent that, despite the fact that large firms suffer the largest breaches, they also -- more frequently -- suffer small ones in which only a fraction of the total information assets are extracted. Assuming that hackers always aim to maximise the volume of exfiltrated information, these small breaches could be considered as only partially successful. More likely, again recognising that organisations are comprised of functional sub-units -- potentially having separate IT systems -- and that attacks occur at the sub-units, then the breach sizes will be related to the subunit sizes. Again recalling the result of \cite{Amaral1998} that the size of the largest sub-unit scales with firm size like $\sim s^{2/3}$, we see an incredible coincidence that the 0.9 quantile of breaches and breach frequency also scale with firm size in this way. This strongly links the risk of breaches to the size of the largest sub-unit. 
  
  This characterization of the risk surface in terms of firm size -- specifically with densities that dictate both how the number of breaches are distributed across firms of different size, and how large breaches tend to be for victimized firms of different size -- allow for the extrapolation of breach statistics onto unobserved populations. For instance, given a population of organizations and their sizes, one can infer the quantity and distribution of breaches that they have suffered. The validity of this most basic extrapolation will involve a ceteris parabis assumption. For instance, one will have to assume that such an organization is similar in attractiveness and permeability to cyber criminals to a publicly traded firm of similar size.
    
  \subsection*{Sector \& Data Breach Risk}
  
   The previous sections focused on identifying universal relationships in data breach risk with a single risk factor, organisation size. Clearly, there are other attributes of organisations that are relevant to data breach risk. We thus consider industrial sector as a risk factor, which may serve as a proxy to identifying relatively homogeneous subpopulations in regard to their frequency of interaction with consumers, and the total volume of personal data that they guard. There are twelve sectors, as defined in~\cite{NASDAQCompanyList}. The sector is likely confounded with firm size, and thus firm size effects. However, due to the limited data, a statistical analysis considering these effects jointly is infeasible. Rather, a simple analysis is done, nuancing the former results.
   
  \begin{figure}[h!]
  \begin{center}
  \centerline{\includegraphics[width=8cm]{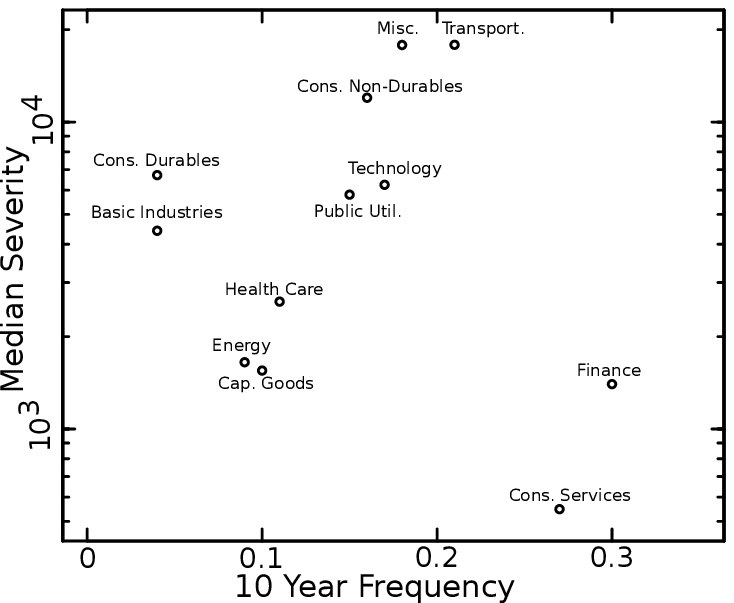}}
  \caption{The median severity and 10 year frequency of breaches for the considered publicly traded firms, grouped into twelve sectors, as defined in~\cite{NASDAQCompanyList}. }
  \label{fig:Sector}
  \end{center}
  \end{figure}
  
Fig.~\ref{fig:Sector} plots the median and 10 year frequency of large breaches for the 12 sectors. 
We propose to rationalise these observations by the hypothesis
that the frequency of breaches is related to the frequency of customer interaction, and that the severity should be related to the volume of personal data guarded by an organisation. Consumer Service, and Finance have small sub-units (i.e., retail shops and bank branches) at which they interact with local customers, and suffer from small but frequent breaches. On the other side, Basic Industries, having large centralized operations, and infrequent customer interaction, implies large yet infrequent breaches. Consumer Durables have a lower breach frequency than Non-Durables, which, by definition, involve more frequent customer contact. Further, the companies with the largest loss in the three categories with highest median loss are Sony (Non-Durables), eBay (Misc.), and UPS (Transportation). These companies all clearly guard large amounts of personal data. Capital Goods (e.g., heavy equipment producers) suffer relatively small breaches, possibly due to a smaller number of customers than retailers of non-durables. These comparisons tend to support the hypotheses posed above.
    
 \section*{Discussion}
 
  Due to the combination of large size and the high potential for immediate financial damage, breaches of personal identities (ids) are among the most disruptive and costly cybersecurity events both for consumers \cite{harrell2013victims} and organisations \cite{sinanaj2014news}. Thus, large breaches translate into immediate financial consequences for organisations, often reflected in drops in stock price \cite{campbell2003economic, garg2003quantifying,acquisti2006there,gatzlaff2010effect}, and reputational damage. Based on this study, the annual total of breached ids is expected to range between half a billion to a billion over the next 5 years. Considering the average cost of more than 200 USD per breached item \cite{ponemon2014}, this translates into hundreds of billions of USD losses per year. Severely worsening the problem is that, not only do data breaches have short term consequences for individuals, but due to sophisticated underground markets for breached data \cite{franklin2007,Rand}, breached personal information is concentrated, enabling efficient subsequent identity fraud \cite{harrell2013victims}. As an illustration, several research studies have established how personal identities may be reconstructed, and individuals re-identified from a few spatio-temporal locations of credit card uses \cite{Montjoyeetal2015} or even from public data \cite{acquisti2009predicting}. Thus, privacy gets increasingly {\it eroded} as the cumulative number of breached ids grows.
  
  Cyber risks, within the general context of the evolving and expanding Internet, are highly dynamic. Being governed by the struggle of IT security technology to keep up with the constant innovation and adaptive nature of cyber crime, ranging from social engineering attacks to the egregious sale of ``zero day'' security vulnerabilities. The struggle is compounded by an ever-growing amount of personal data stored online, and a growing attack surface due to the adoption of mobile computing paradigms. Due to the clear potential for damage, it is crucial that the risk of data breaches be well understood. Despite the dynamic context of data breaches, and a relatively short history, we have specified a statistical model for risk that successfully unearths clear statistical regularities, allowing for novel understanding of this risk. 
   
  The frequency of large breaches (having in excess of fifty thousand ids) was found to be stable since 2007, with a rate of 76 events per year. Despite the relatively low rate, since 2007 nearly 2 billion pieces of personal information have been breached, hinting at the extreme size of large breaches. For breach sizes, we considered three possibilities: (i) there is a maximum possible breach size, (ii) this maximum has grown, and (iii) large breaches became larger over time. These statements were formally tested on the data (from January $1^{st}$, 2007 until April $15^{th}$, 2015). By Extreme Value Theory (EVT), it was found that a highly significant maximum breach size exists, and is growing with time like $t^{0.84}$, where the current maximum breach size is about 200 million, and is expected to grow fifty percent over the next five years. This feature is highly relevant for policymakers and re-insurance companies who are concerned with evaluating and managing the total risk.
  
  On a critical note, this upper limit may not be entirely valid. That is, EVT cannot consider the presense of an unobserved regime beyond the {\it usual} df. Such singular extreme statistics, also called {\it Dragon-Kings}, have been observed in, e.g., city sizes, financial crashes, and nuclear accidents \cite{sornette2012dragon,WheatleyNuclear2015}. For example, Facebook gathers personal data from 1.4 billion users (as of January 2015, and not counting Whatsapp (700 million users) acquired by Facebook in 2014), and the National Security Agency (NSA) most probably gathers personal information about several billion people worldwide. Though these organisations certainly operate with incomparably more resources, and at a much high level of information security, the chance of massive attacks on personal data gathered and stored by Internet giants and governments cannot be excluded in the future -- especially considering the size effect by which large organisations are more frequently attacked, discussed below.
  
  The df of breach sizes was found to be well modeled by a Pareto (Power law) df with a linearly shrinking shape parameter, and maximum breach size given by the EVT estimate. The estimate of this model, having shape parameter $0.57~(0.05)$ in 2007, is expected to have become much heavier tailed with shape parameter $0.37$ in 2015. Under this current model, given that a breach is in excess of fifty thousand ids, there is a ten percent chance that the breach exceeds ten million ids.
      
  Next, the connection between organisation size, measured by market capitalization, and the risk of data breaches was unearthed. It was found that the frequency of breaches scales with organisation size like $\sim s^{0.6}$, indicating that larger organisations are victimised at much higher frequencies than their smaller counterparts. Further, the largest breach sizes (quantified by the 0.9 quantile) were found to scale with organisation size $\sim s^{0.66}$ for firms with market capitalisation between one million and ten billion US\$. Above ten billion US\$, the scaling reached a plateau. Thus we identified that the effective \emph{risk surface} scales with organisation size with at exponent around 0.6. This can be thought of as a fractal scaling relationship. Alternatively, recognising that organisations are comprised of functional sub-units -- potentially having separate IT systems -- and that attacks occur at the sub-units, then the breach sizes will be related to the subunit sizes. In \cite{Amaral1998}, it was found that the size of the largest sub-unit scales with firm size like $\sim s^{2/3}$. That this scaling relationship is similar to that of the risk surface, is highly suggestive that the size of the largest sub-unit -- potentially housing the main IT system and data -- defines the risk level. This is a precious first step towards establishing a relationship between data breaches and the underlying structure of organisations in which they tend to occur.
  
  As the damage of data breaches is a cumulative erosion process, we also studied the cumulative sum of data breaches. Like many negative externalities in the economy \cite{tybout1972pricing}, the phenomenon of privacy erosion is hard to measure. For instance, the cumulative sum likely overestimates the erosion as, to some extent, the exfiltrated data degrades over time. In particular, users can change their passwords and cancel their credit cards. However, other aspects of the victims identity -- such as name, address, and social security number are more persistent. Thus, in reality, the true erosion is somewhere in between an instantaneous and cumulative process. However, this approach provides a transparent quantification upon which discussions of past and future risk may be had. To date, the total amount of breached personal data, well approximated by the sum of large breaches, is around 2 billion personal id items. Although some individuals may have been subjected to multiple breaches, the amount of aggregate loss is considerable, e.g., compared to the number of Internet users (3.5 billion) or even to the world population (7 billion). Projections based on the best models suggest that the growth of the cumulative sum is accelerating, and is expected to double (surpassing the number of Internet users) in the next 5 years. Finally, it is important to note that about 30 percent of events have unknown breach size, and thus the statistics above are underestimates.
  
  Our results provide detailed insights on the aggregate statistics of data breaches, their impact on organisations, as well as on the long-term effects for consumers and society. In fact, the provided estimates of the risk level are conservative given the fact that we are only considering the subset of data breaches that have been discovered, reported, and submitted to the online databases. Our findings suggest that people should not only worry about privacy erosion the way we experience it nowadays, by assuming that our personal data will remain confined and exploited only by ``trustworthy'' private organisations and governmental agencies. On the contrary, people should expect that the personal data they have uploaded on online platforms, such as social networks, may be either suddenly or progressively disseminated, first in underground markets, and then publicly.   
  
Finally, this work supports and encourages further emphasis on addressing cyber risks at both organisational, and governmental levels. This risk should not be thought of in terms of representative or typical data breaches, which is a completely inadequate concept given the heavy-tailed nature of the pdfs. Rather one must consider the full distribution. Moreover, this is a system where the total risk is dominated by the few extremes, which are inherently stochastic. Thus, policies should be adapted to this regime where standard actuarial methods, insurance by the mean, do not apply. In terms of defence, this also calls for measures aimed at stopping the big events, the cascades, perhaps by decentralizing IT systems in big organisations.

\bibliographystyle{abbrv}
\bibliography{cyber}

\section*{Acknowledgements}
T.M. acknowledges support by the Swiss National Science Foundation (award PA00P2-145368) and the National Science Foundation under award CCF-0424422 (TRUST).

S.W. acknowledges input from Travis Biehn.

\section*{Author contributions statement}
S.W. designed the analysis with input from D.S. and T.M., S.W. conducted the analysis with data from T.M., all authors contributed to the writing of the paper. All authors reviewed the manuscript. 

\section*{Additional information}
The authors declare no competing financial interests.

\end{document}